\documentclass[a4paper,twocolumn,english,prl,superscriptaddress,longbibliography,fixfloat,notitlepage]{revtex4-2}
\pdfoutput=1
\usepackage[T1]{fontenc}
\usepackage[latin9]{inputenc}
\usepackage{xcolor}
\usepackage{pdfcolmk}
\usepackage{float}
\usepackage{amsmath}
\usepackage{amssymb}
\usepackage{graphicx}
\usepackage{wasysym}
\PassOptionsToPackage{normalem}{ulem}
\usepackage{ulem}

\makeatletter

\pdfpageheight\paperheight
\pdfpagewidth\paperwidth

\providecolor{lyxadded}{rgb}{0,0,1}
\providecolor{lyxdeleted}{rgb}{1,0,0}
\DeclareRobustCommand{\lyxadded}[3]{{\color{lyxadded}{}#3}}
\DeclareRobustCommand{\lyxdeleted}[3]{{\color{lyxdeleted}\lyxsout{#3}}}
\DeclareRobustCommand{\lyxsout}[1]{\ifx\\#1\else\sout{#1}\fi}

\usepackage{xcolor}
\definecolor{darkblue}{rgb}{0.1,0.2,0.6} 
\definecolor{lightblue}{rgb}{0.1,0.1,1.0}
\definecolor{darkred}{rgb}{0.8,0.1,0.2}
\usepackage[colorlinks,citecolor=lightblue,linkcolor=darkblue,urlcolor=lightblue] {hyperref}

\newcommand{\beginsupplement}{%
\pagebreak
\widetext
\setcounter{equation}{0}
\setcounter{figure}{0}
\setcounter{table}{0}
\setcounter{page}{1}
\makeatletter
\renewcommand{\theequation}{S\arabic{equation}}
\renewcommand{\thefigure}{S\arabic{figure}}
}%

\makeatother

\usepackage{babel}
\begin{document}
\global\long\def\E{\mathrm{e}}%
\global\long\def\D{\mathrm{d}}%
\global\long\def\I{\mathrm{i}}%
\global\long\def\mat#1{\mathsf{#1}}%
\global\long\def\vec#1{\mathsf{#1}}%
\global\long\def\cf{\textit{cf.}}%
\global\long\def\ie{\textit{i.e.}}%
\global\long\def\eg{\textit{e.g.}}%
\global\long\def\vs{\textit{vs.}}%
 
\global\long\def\ket#1{\left|#1\right\rangle }%

\global\long\def\etal{\textit{et al.}}%
\global\long\def\tr{\text{Tr}\,}%
 
\global\long\def\im{\text{Im}\,}%
 
\global\long\def\re{\text{Re}\,}%
 
\global\long\def\bra#1{\left\langle #1\right|}%
 
\global\long\def\braket#1#2{\left.\left\langle #1\right|#2\right\rangle }%
 
\global\long\def\obracket#1#2#3{\left\langle #1\right|#2\left|#3\right\rangle }%
 
\global\long\def\proj#1#2{\left.\left.\left|#1\right\rangle \right\langle #2\right|}%
\global\long\def\mds#1{\mathds{#1}}%

\title{{\Large{}Transport in Stark Many Body Localized Systems}}
\author{Guy Zisling}
\affiliation{Department of Physics, Ben-Gurion University of the Negev, Beer-Sheva
84105, Israel}
\author{Dante M. Kennes}
\affiliation{Institute for Theory of Statistical Physics, RWTH Aachen University,
and JARA Fundamentals of Future Information Technology, 52056 Aachen,
Germany}
\affiliation{Max Planck Institute for the Structure and Dynamics of Matter, Center
for Free Electron Laser Science, Luruper Chaussee 149, 22761 Hamburg,
Germany}
\author{Yevgeny Bar Lev}
\affiliation{Department of Physics, Ben-Gurion University of the Negev, Beer-Sheva
84105, Israel}
\email{ybarlev@bgu.ac.il}

\begin{abstract}
{\normalsize{}Using numerically exact methods we study transport in
an interacting spin chain which for sufficiently strong spatially
constant electric field is expected to experience Stark many-body
localization. We show that starting from a generic initial state,
a spin-excitation remains localized only up to a finite delocalization
time, which depends exponentially on the size of the system and the
strength of the electric field. This suggests that bona fide Stark
many-body localization occurs only in the thermodynamic limit. We
also demonstrate that the transient localization in a finite system
and for electric fields stronger than the interaction strength can
be well approximated by a Magnus expansion up-to times which grow
with the electric field strength.}{\normalsize\par}
\end{abstract}
\maketitle
\emph{Introduction}.---Statistical mechanics assumes that isolated,
interacting systems with many degrees of freedom always approach the
state of thermal equilibrium. More than a decade ago, it was argued
that in the presence of a sufficiently strong disorder, this assumption
can be defied, using a mechanism known as many-body localization (MBL)
\citep{Basko2006a,Gornyi2005,Altman2014,Nandkishore2014,Vasseur2016,Abanin2017,Abanin2019}.
If such systems are isolated from the environment they will never
thermalize. Perfect isolation from the environment is challenging
in conventional condensed matter systems due to inevitable presence
of phonons \citep{Basko2007a,Ovadia2014}, however evidence of MBL
was obtained in numerous experiments in cold atoms in both one-dimensional
\citep{Schreiber2015a,Bordia2015,Smith2015} and two-dimensional systems
\citep{Choi2016}. While coupling to an external environment or a
noise source is detrimental to MBL \citep{Znidaric2010a,Gopalakrishnan2016a},
it was shown to be stable to periodic driving at sufficiently high
frequencies. A phenomenon known as Floquet--MBL \citep{Lazarides2014,Ponte2014,Abanin2014,Bairey2017}. 

\begin{figure}[t]
\includegraphics[width=8.6cm]{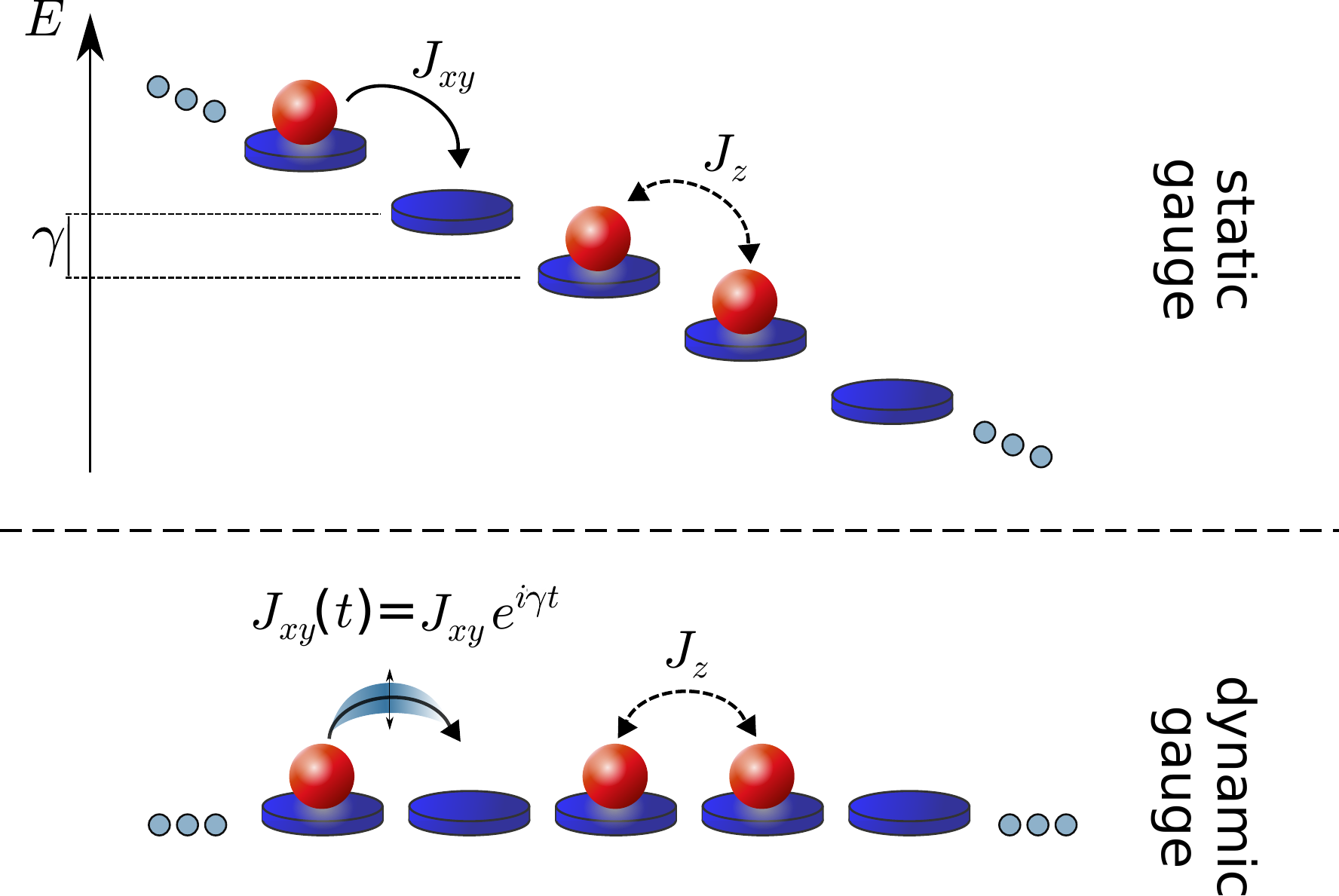}\caption{\label{fig:Artistic-Figure}A schematic representation of the Stark
localization problem in two gauges. The upper panel shows the static
gauge, where particles are subject to a tilted potential. The bottom
panel shows a dynamic gauge, where the scalar potential is written
as a ``vector potential,'' which produces time-dependent hopping.}
\end{figure}
Theoretical arguments in favor of MBL require the localization of
\emph{all} the single-particle states \citep{Basko2006a,Imbrie2014}.
For quenched disorder this requirement is naturally satisfied in one
and two-dimensional systems due to Anderson localization \citep{Anderson1958b}.
Various attempts to relax this requirement were performed by considering
models where some of the single-particle states are delocalized \citep{Nandkishore2014c,Li2015,Li2016,Modak,Modak2015,Hyatt2016,Lev2016,Vasseur2015b},
as also translationally invariant models where all of the states are
delocalized in the absence of interactions \citep{Carleo2012,Schiulaz2013,Grover2013,Schiulaz2014,Yao2014,Hickey2014,Papic2015,VanHorssen2015,Pino2015,Lev2016}.
However, the observed localization is far from being convincing and
typically suffers from severe finite-size effects \citep{Papic2015}.
Moreover, while some of these models show robust localization for
special initial states, most initial states are apparently delocalized
\citep{Carleo2012,Lev2016}.

Anderson localization is not the only mechanism which can be used
to localize the single-particle states. Single-particle states can
be localized by a periodic-in-time, spatially uniform electric field
at certain drive frequencies \citep{Dunlap1986,Dunlap1988}, and also
by a static uniform electric field and any field strength \citep{Wannier1960}.
The former is known as dynamic localization, and the later as Wannier-Stark
localization. While it was shown that dynamic localization is not
stable to the addition of interactions \citep{Luitz2017b}, Wannier-Stark
localization was argued to be stable to interactions for sufficiently
strong electric fields \citep{Schulz2019,VanNieuwenburg2019}, a phenomenon
dubbed Stark-MBL. Via a gauge change, constant electric field can
be replaced by a time-dependent vector potential (see Fig\@.~\ref{fig:Artistic-Figure}).
Therefore the Stark problem is equivalent to a periodically driven
translationally invariant interacting model; see Eq. (\ref{eq:time-dep-ham}).
The mechanism behind Stark-MBL is currently under debate, since many
of the arguments of Refs.~\citep{Basko2006a,Imbrie2014} cannot be
readily applied due to proliferation of resonances, which are known
to induce asymptotic delocalization in certain cases \citep{Michailidis2017}.
It was proposed that Stark-MBL follows from an approximate ``shattering''
of the Hilbert space due to an almost-conservation of the dipole moment
\citep{VanNieuwenburg2019,Pai2018,Khemani2020}. This argument is
however applicable only for an infinite electric field, $\gamma$,
where jumps between sites are prohibited due to energy conservation
(see Fig\@.~\ref{fig:Artistic-Figure}), and cannot be easily generalized
for finite and modest electric fields where the Stark-MBL transition
ostensibly occurs \citep{Schulz2019,VanNieuwenburg2019,SuppMat2021}.

The dynamics in both localized and delocalized phases was studied
theoretically \citep{Schulz2019,VanNieuwenburg2019,Zhang2020,Doggen2021}
and experimentally \citep{Guardado-Sanchez2020,Morong2021,Scherg2021}
starting from special initial states. Two-dimensional systems are
delocalized and show subdiffusive transport \citep{Guardado-Sanchez2020,Zhang2020}.
For one-dimensional systems and sufficiently strong electric fields
both charge-density wave (CDW) \citep{Schulz2019,VanNieuwenburg2019,Morong2021,Scherg2021}
and domain-walls initial states \citep{Doggen2021} do not appear
to melt completely. In fact in Ref.~\citep{Doggen2021} it was argued
that the system is localized in the thermodynamic limit, for any nonzero
electric field, though Ref.~\citep{Yao2021} suggested that this
is a special property of domain-wall initial states.

In this Letter, we consider the nonequilibrium dynamics in a one-dimensional
Stark-MBL system starting from a \emph{generic} initial state, which
corresponds to an average over \emph{all} possible initial states.
We demonstrate that in both presumably delocalized, and localized
regions, a local spin excitation remains localized for increasingly
long times when the system size is increased, suggesting that transport
might be completely suppressed only in the thermodynamic limit.

\emph{Model}.---The interacting Stark model is described by the following
Hamiltonian,

\begin{align}
\hat{H} & =\sum_{j=1}^{L-1}\frac{J_{xy}}{2}\left(\hat{S}_{j}^{+}\hat{S}_{j+1}^{-}+\text{h.c.}\right)+J_{z}\hat{S}_{j}^{z}\hat{S}_{j+1}^{z}+\sum_{j=1}^{L}W_{j}\hat{S}_{j}^{z},\label{eq:hamiltonian_spins}
\end{align}
where $L$ is the length of the spin-chain, ``h.c.'' denotes the
hermitian conjugate, $\hat{S}_{j}^{\pm}$, $\hat{S}_{j}^{z}$ are
spin-1/2 operators, $J_{xy}$ is the strength of the flip-flop term,
$J_{z}$ is the strength of the Ising term, and $W_{j}=\left(\gamma j+\alpha j^{2}/L^{2}\right)$
is a spatially varying potential, where $\gamma$ corresponds to an
electric field, $\alpha/L^{2}$ is the magnitude of a shallow parabolic
trap that we add in order to break some of the symmetries of the system,
following Ref.~\citep{Schulz2019}. The system conserves the total
magnetization, $\hat{M}=\sum_{j}\hat{S}_{j}^{z}$ and in the thermodynamic
limit is translationally invariant (for $\alpha=0$). Through this
work we use open boundary conditions and set $J_{xy}=2$, $J_{z}=1$
and $\alpha=0.5$, verifying that our results do not change qualitatively
for other $\alpha$s and $J_{z}$s, as also boundary conditions (see
\citep{SuppMat2021}). Via the Jordan-Wigner transformation \citep{Jordan1928},
the model is equivalent to a system of spinless interacting fermions
moving in a uniform electric field, however for the clarity of the
presentation we proceed using the spin formalism.

A number of works show an apparent ergodicity breaking for $\gamma\gtrsim1.5$
\citep{Schulz2019,VanNieuwenburg2019} (see also \citep{SuppMat2021}).
In this Letter, using two numerically exact methods we study spin-transport
in this model.

\emph{Methods}.---
\begin{figure}[t]
\includegraphics{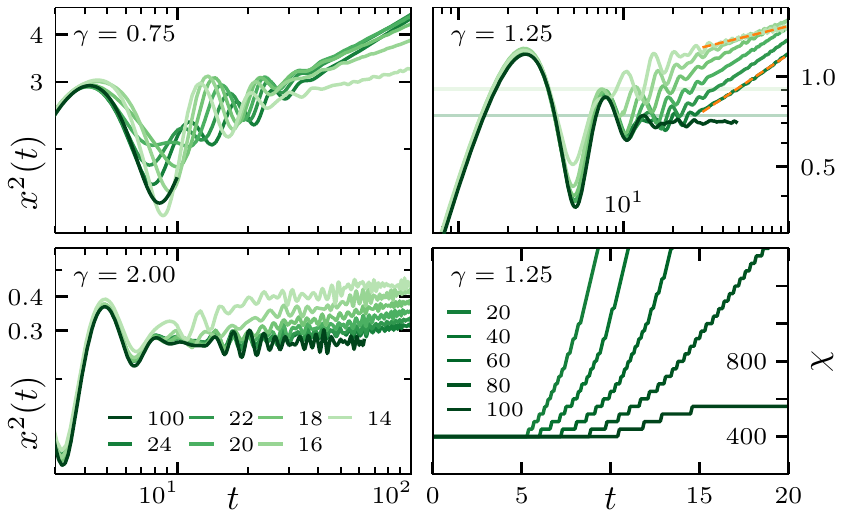}\caption{\label{fig:Mean-square-displacement}Mean-square displacement (MSD)
as a function of time (top panels and bottom left) for $L\in[14,24]$
(Krylov based method) and $L=100$ (tDMRG). The orange dotted line
correspond to power-law fits, while the horizontal lines indicate
the plateau of the MSD calculated by taking the mean of the MSD between
the 2nd and the 3rd peaks. The color of the plateau lines matches
the coloring of the corresponding system size. \emph{Bottom right}.
Bond dimension, $\chi$ as a function of time obtained using tDMRG
for $L\in[20,100]$ and with a fixed discarded weight $10^{-7}$.
All plots were obtained for $J_{xy}=2,\ J_{z}=1$ and $\gamma=1.25$.}
\end{figure}
To assess spin-transport in the system for various electric fields,
$\gamma$, we calculate the infinite temperature spin-spin correlation
function,
\begin{equation}
G_{n}\left(t\right)=\frac{1}{\mathcal{N}}\textrm{Tr}\left[\hat{S}_{n}^{z}\left(t\right)\hat{S}_{L/2}^{z}\right],\label{eq:correlation}
\end{equation}
where $\mathcal{N}$ is the Hilbert space dimension, and $\hat{S}_{n}^{z}\left(t\right)$
is the Heisenberg evolution of $\hat{S}_{n}^{z}$. This correlation
function describes the spatial spreading of an initially local spin
excitation on top of an infinite temperature state. The squared width
of the excitation, is given by, $x^{2}\left(t\right)=\sum_{n}n^{2}\left(G_{n}\left(t\right)-G_{n}\left(0\right)\right)$
and is analogous to the mean-square displacement (MSD). For diffusive
transport, $x^{2}\sim2Dt$, with $D$ coinciding with the diffusion
coefficient calculated from the corresponding Kubo formula \citep{Steinigeweg2009a,Yan2015,Luitz2016,Steinigeweg2017}.

We compute $G_{n}\left(t\right)$ using two complementary numerically
exact methods. In the first method we work at a zero magnetization
sector, with the Hilbert space dimension $\mathcal{N}=\binom{L}{L/2}$
and utilize dynamical typicality to reduce the trace in Eq.~(\ref{eq:correlation})
to a unitary propagation of a random initial state taken from the
Haar distribution \citep{Popescu2006,Luitz2016}. We then average
over a small number of such samples. Our initial condition therefore
corresponds to a generic initial state with volume law entanglement.
We would like to stress that while the generation of such a highly
entangled pure state is probably close to impossible experimentally,
we could equally well take a random product state, which \emph{can}
be realized experimentally. Such a state would produce an equivalent
result, though it would require more averaging over the initial states
to to sample the correlation function in Eq.~(\ref{eq:correlation}).

The unitary evolution is performed using a Krylov subspace method
\citep{Moler2003}. Given the exponential scaling of the Hilbert space
dimension we are able access system sizes of $L\lesssim24$, which
correspond to $\mathcal{N}\leq2\,704\,156$, though we can propagate
the system for quite long times. As a complimentary method, which
provides us access to large systems sizes, we use the time-dependent
density matrix renormalization group (tDMRG) \citep{Schollwock2011}.
In this method the wavefunction is represented as a matrix product
state (MPS), built of matrices with a maximal dimension $\chi$, called
the bond-dimension. The bond-dimension sets the maximum entanglement
that the MPS can accommodate. If the bond dimension is set to be smaller
than $\chi<d^{L/2}$, where $d$ is the local Hilbert space dimension,
the error in the MPS representation of the wavefunction is bounded
by the truncation weight. In our simulations, we set the truncation
weight to $10^{-7}$ allowing the bond dimension to grow dynamically
during the propagation. Since for ergodic systems the entanglement
is typically increasing linearly with time, the computational effort
increases exponentially. We checked for converges of our results by
decreasing the truncation weight to $10^{-8}$. In tDMRG we use \emph{all}
magnetization sectors, and obtain $G_{n}\left(t\right)$ by the Heisenberg
evolution of $\hat{S}_{n}^{z}\left(t\right)$, using the computational
method detailed in Ref.~\citep{kennes2016extending}. Due to the
equivalence between the ensembles of fixed and varying magnetization,
in the thermodynamic limit, both Krylov based and tDMRG results are
expected to agree up to some finite time when the finite size effects
become important.

\emph{Results}.---We calculate the MSD for a number of electric fields,
$\gamma=0.75-3$ and various system sizes $L=14-24$ using the Krylov
subspace method, and for sizes $L=20-100$ using tDMRG. The results
are presented in Fig.~\ref{fig:Mean-square-displacement}. For times
$t\leq t^{\star}\left(\gamma,L\right)$ an initial growth of the MSD
is followed by a localization plateau. This plateau is visible for
$\gamma\apprge1$, and becomes even more pronounced for larger system
sizes. For all the studied $\gamma$'s, including a regime where according
to Refs.~\citep{Schulz2019,VanNieuwenburg2019} (see also \citep{SuppMat2021})
the system is expected to be strongly localized, the late-time dynamics
of a \emph{finite} system is always delocalized, which allows us to
identify the time $t^{\star}\left(\gamma,L\right)$, as the delocalization
time. Note that our data suggests, that for $\gamma\apprge1$ the
system becomes localized only in the thermodynamic limit. The observed,
apparently subdiffusive growth of the MSD for $t>t^{\star}\left(\gamma,L\right)$,
which is consistent with previous experimental \citep{Guardado-Sanchez2020}
and theoretical works \citep{Zhang2020a,Gromov2020,Feldmeier2020,Yao2021},
is therefore a finite-size effect, and will not be considered further
in this Letter (see however \citep{SuppMat2021}). For $\gamma\leq1$
our results are not conclusive, since the delocalization time, if
it exists here, is very short, and the plateau in the MSD is not clearly
visible. But, we do see that for $\gamma=0.75$ the fast growth of
the MSD is pushed to later times for larger system sizes, which hints
that localization at the \emph{thermodynamic limit} might occur for
all $\gamma>0$. A similar suggestion was recently raised in Ref.~\citep{Doggen2021}.

\begin{figure}[t]
\includegraphics[width=1\columnwidth]{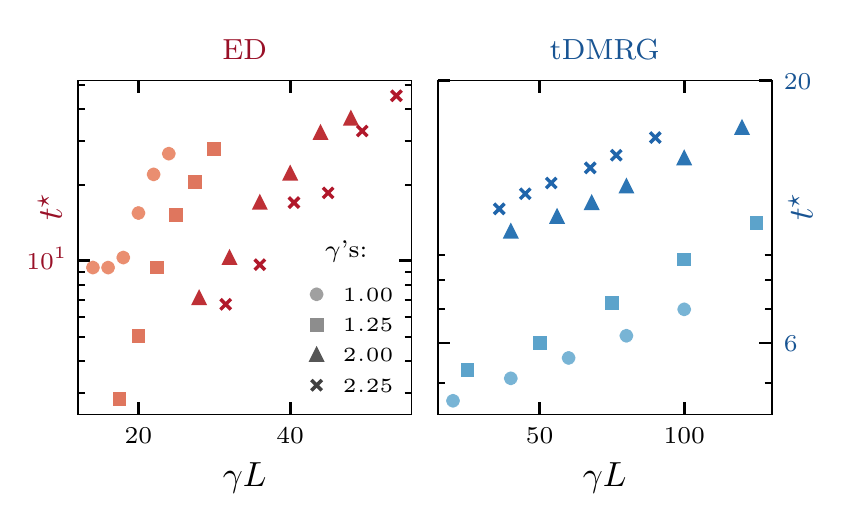}\caption{\label{fig:t-star-scaling}Delocalization time $t^{\star}$ as a function
of $\gamma L$ for $\gamma=1,1.25,2,2.25$, as extracted from Krylov
based method (left panel, $L\in[14,24]$) and tDMRG (right panel,
$L\in[20,100]$). For all data points $J_{z}=1$.}
\end{figure}
The localization--delocalization transition at a finite time, $t^{\star}\left(\gamma,L\right)$,
can also be seen from the growth of the bond-dimension in tDMRG to
maintain a chosen accuracy of the results (discarded weight). For
$t\leq t^{\star}\left(\gamma,L\right)$ a modest bond-dimension is
required, while for $t>t^{\star}\left(\gamma,L\right)$ to keep the
same accuracy of the numerical evolution an increasingly larger bond-dimension
is required. We stress that the bond-dimension is not a physical quantity
and we only use it as an indicator of delocalization to obtain, $t^{\star}\left(\gamma,L\right)$\footnote{The entanglement entropy here would be an entanglement of an MPO,
which is also not a measurable physical quantity}.

To quantitatively study the dependence of $t^{\star}\left(\gamma,L\right)$
on $\gamma$ and $L$, we extract it using two independent methods.
For the Krylov subspace method it is extracted from the intersection
point between two straight lines on a log-log scale: the plateau of
the MSD (see caption in Fig.~\ref{fig:Mean-square-displacement})
and the apparent subdiffusive growth (dashed orange lines in Fig.~\ref{fig:Mean-square-displacement}).
For tDMRG we define $t^{\star}\left(\gamma,L\right)$ as the time
when the bond-dimension departs from its initial value (set to 400).
While these definitions are of-course arbitrary, using different definitions
did not result in a qualitative change. In Fig.~\ref{fig:t-star-scaling}
we show the delocalization time, $t^{\star}\left(\gamma,L\right)$
plotted vs $\gamma L$ on a semi-log scale for various tilts of the
potential $\gamma$. We find that both Krylov subspace and tDMRG methods,
suggest that the delocalization time increases exponentially both
with $\gamma$ and $L$, namely $t^{\star}\sim\exp\left[\gamma L\right]$,
such that true localization is obtained only in the thermodynamic
limit. Remarkably, the tDMRG simulation of this system becomes \emph{easier},
namely with the same computational resources for \emph{larger} system
sizes one can go to \emph{longer} times. This indicates a change in
the \emph{bulk} dynamics, when the size of the system is increased,
even though the Hamiltonian is local.

\begin{figure}[t]
\includegraphics{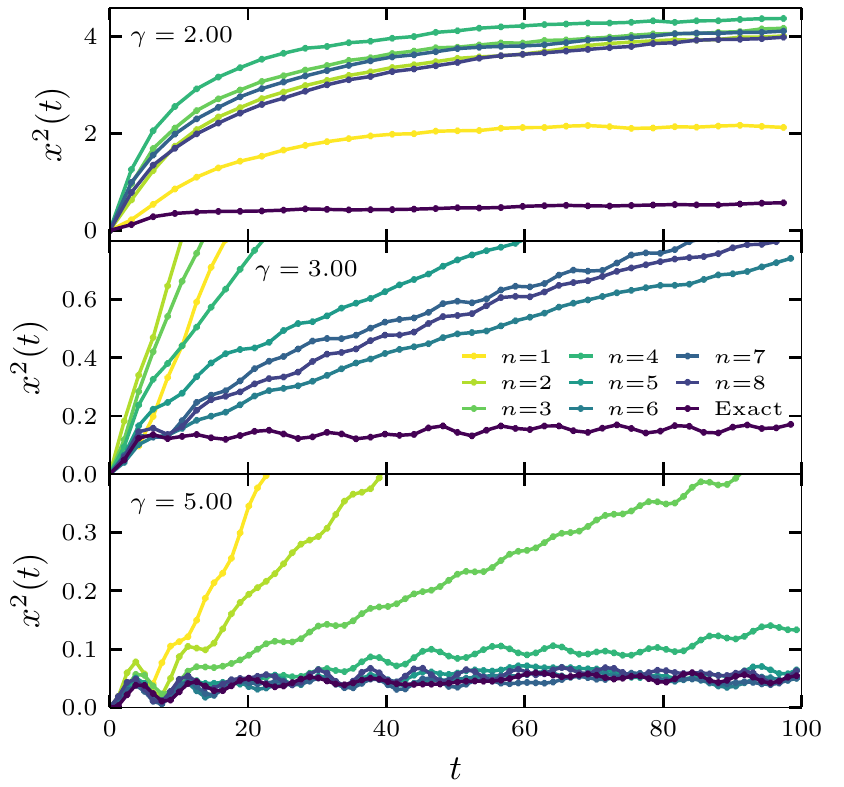}\caption{\label{fig:Magnus}Mean-squared displacement as a function of time
for various electric fields. The darkest lines correspond to numerically
exact results obtained by using Eq.~(\ref{eq:time-dep-ham}) for
propagation. The colored lines with increasing brightness corresponds
to evolution using effective Hamiltonians Eq.~(\ref{eq:effective_hamiltonian_def}),
obtained from a truncated Magnus expansion. For all panels, $J_{xy}=2,\ J_{z}=1,\ L=14$.}
\end{figure}
\emph{Magnus expansion}.---In order to better understand the dependence
of $t^{\star}\left(\gamma,L\right)$ on $\gamma$ we apply a time-dependent
unitary transformation $\hat{U}\left(t\right)\equiv e^{-i\gamma t\sum_{j}j\hat{S}_{j}^{z}}$
to Eq.~(\ref{eq:hamiltonian_spins}), which corresponds to a gauge
change, replacing the potential term in Eq.~(\ref{eq:hamiltonian_spins})
by a time-dependent ``vector potential.'' This yields the following
time-dependent Hamiltonian,

\begin{equation}
\hat{H}\left(t\right)=\sum_{j=1}^{L-1}\left[\frac{J_{xy}}{2}\left(e^{-i\gamma t}\hat{S}_{j}^{+}\hat{S}_{j+1}^{-}+\text{h.c.}\right)+J_{z}\hat{S}_{j}^{z}\hat{S}_{j+1}^{z}\right],\label{eq:time-dep-ham}
\end{equation}
where the electric field, $\gamma$, takes the role of a frequency.
The static part of the Hamiltonian is trivially localized and has
a spectrum composed of highly degenerate energy bands, which differ
by a number of domain walls. It takes an energy of $J_{z}/4$ to annihilate
or create a domain wall, and therefore the bands are equally spaced.
The time-dependent hopping facilitates transport in the system by
two possible processes: either by connecting the various bands, or
by higher order, virtual transitions from some state in a band to
a different state in the same band. For $\gamma\gg J_{z}/4$ both
processes are suppressed since multiple spin rearrangements are required
to absorb the energy of the ``photon'' and the system is expected
to be in a long-lived prethermal state described by the time-averaged
Hamiltonian (which here coincides with the static part of $\hat{H}\left(t\right)$)
up to times $t^{\star}\sim\exp\left[\gamma/J_{z}\right]$ \citep{Abanin2015a,Abanin2015c,Abanin2015d}.
A slightly different scaling was suggested in Ref.~\citep{Khemani2020}.
We have checked that for larger $J_{z}$, the apparent localization--delocalization
transition shifted to larger $\gamma$s \citep{SuppMat2021}.

The stroboscopic evolution of the system is determined by an effective
Hamiltonian, which is defined from the one-period propagator,
\begin{equation}
\hat{U}\left(T\right)=\E^{-i\hat{H}_{\text{eff}}T}=\mathcal{T}\exp\left[-i\int_{0}^{T}\mathrm{d}\bar{t}\hat{H}\left(\bar{t}\right)\right],\label{eq:effective_hamiltonian_def}
\end{equation}
where $\mathcal{T}$ corresponds to time-ordering, and $T=2\pi/\gamma$
is the period. For $\gamma\gg J_{z}/4$ we can approximate $\hat{H}_{\text{eff}}$
by a Magnus expansion in $\gamma^{-1}$ \citep{Blanes2009}. For $\gamma$
smaller than the many-body band-width, this expansion is \emph{not}
guaranteed to converge, but it can approximate the dynamics of the
system up to some optimal order \citep{Mori2016}. We use a recursive
formula described in Ref.~\citep{Blanes2009} to obtain $\hat{H}_{\text{eff}}$
up to order $n=10$ for $L=14$. Fig.~\ref{fig:Magnus} shows the
stroboscopic evolution of the MSD computed numerically using $\hat{H}_{\text{eff}}^{\left(n\right)}$,
which is $\hat{H}_{\text{eff}}$ truncated to an order $n$. We see
that for $\gamma\leq2$ the Magnus expansion fails to approximate
the dynamics even for short times, while for $\gamma=3,5$ as the
Magnus order $n$ increases, the approximate solution approaches the
exact solution for longer times (it is hard to reliably extract $t_{\text{magnus}}$
from our data to obtain the functional dependence on $n$, but see
Ref.~\citep{Kuwahara2016b}). There is little to no dependence of
$t_{\text{magnus}}$ on the system size (see \citep{SuppMat2021}).
Interestingly, the long times dynamics of $\hat{H}_{\text{eff}}^{\left(n\right)}$
is diffusive with a diffusion coefficient which decreases with $n$
\citep{SuppMat2021}, even for $\gamma=5$, where the system is expected
to be strongly localized \citep{Schulz2019,VanNieuwenburg2019}.

\emph{Discussion}.---In this Letter, using two complementary numerically
exact methods, we have examined the dynamics of a spin-excitation
starting from a generic initial condition in a spin-chain which is
expected to exhibit Stark-MBL. For $\gamma>J_{z}$ we find strong
evidence of a finite delocalization time, $t^{\star}\left(L,\gamma\right)$,
which scales exponentially with both the size of the system and the
electric field, namely $t^{\star}\left(L,\gamma\right)\sim\exp\left[\gamma L/J_{z}\right]$.
For intermittent times $t<t^{\star}$ the spin-excitation is localized,
while for $t>t^{\star}$ it delocalizes in a manner consistent with
subdiffusion \citep{Guardado-Sanchez2020}. This strongly suggests
that for $\gamma\apprge J_{z}$, Stark-MBL strictly occurs only in
the thermodynamic limit, $L\to\infty$, while any \emph{finite} system
is ultimately delocalized for sufficiently long times. For $\gamma\leq J_{z}$
and system sizes and times accessible to us, the localization regime
is not apparent. Nevertheless, we do see that the dynamics is delayed
with increasing the system size, which can be consistent with a localization
length larger than the system size $\xi\left(\gamma\right)\gg L$.
It is therefore plausible to conjecture that that Stark-MBL in the
thermodynamic limit occurs for all $\gamma>0$, which is consistent
with the conjecture in Ref.~\citep{Doggen2021}.

In the dynamic gauge, where the electric field is replaced by a periodically
driven flip-flop term such that $\gamma$ plays the role of the frequency,
it is rigorously known that for $\gamma\gg J_{z}$ the heating time
is exponential in $\gamma/J_{z}$ \citep{Abanin2015a,Abanin2015c,Abanin2015d}.
We show that for sufficiently large electric fields, up to time $t_{\text{magnus}}$,
the dynamics is well approximated by a \emph{static} effective Hamiltonian
obtained from a Magnus expansion truncated up to order $n$. This
time increases with both $\gamma/J_{z}$ and $n$ (cf. Ref.~\citep{Kuwahara2016b}).
The first order of the expansion is given by $\hat{H}_{\text{eff}}^{\left(1\right)}=J_{z}\sum_{j=1}^{L-1}\hat{S}_{j}^{z}\hat{S}_{j+1}^{z}+O\left(J_{z}/\gamma\right)$.
The spectrum of $\hat{H}_{\text{eff}}^{\left(n\right)}$ is composed
of equally space bands, $J_{z}/4$ distance apart, with a bandwidth
of $O\left(J_{z}/\gamma\right)$ \citep{SuppMat2021}. Therefore,
for $\hat{H}_{\text{eff}}^{\left(n\right)}$ the situation is similar
to models of quasi-MBL, which show asymptotic delocalization \citep{Schiulaz2014,Papic2015,Michailidis2017}.
Indeed all $\hat{H}_{\text{eff}}^{\left(n>1\right)}$ show diffusion
at long times, with a diffusion constant decreasing with $n$ \citep{SuppMat2021}.
We would like to stress that the delocalization of $\hat{H}_{\text{eff}}^{\left(n\right)}$
occurs \emph{before} the delocalization in Eq.~(\ref{eq:hamiltonian_spins})
and Eq.~(\ref{eq:time-dep-ham}) at time $t^{\star}$, and therefore
Magnus expansion does \emph{not} capture the delocalization regime
of Eq.~(\ref{eq:hamiltonian_spins}) and Eq.~(\ref{eq:time-dep-ham}).
It does suggest that the localization mechanism of Stark-MBL is probably
different from Floquet-MBL, where the effective Hamiltonian is expected
to be non-ergodic \citep{Lazarides2014,Abanin2014}.

While the analysis we provided explains the transient localization
regime, it does \emph{not} explain why the delocalization time increases
with the size of the system, suggesting that Stark-MBL happens only
in the thermodynamic limit. This conclusion remains qualitatively
robust for both open and periodic boundary conditions, in the static
and dynamic gauges, and with and without the parabolic potential in
Eq. (\ref{eq:hamiltonian_spins}) \citep{SuppMat2021}. A possible
explanation could be that the measure of delocalized states is vanishing
in the thermodynamic limit. This would also explain why localization
appears to be robust for CDW and domain-wall initial states. We leave
the exploration of this avenue to future studies.
\begin{acknowledgments}
We would like to thank Achilleas Lazarides for insightful discussions
and constructive comments on the manuscript, and Tomotaka Kuwahara
for providing technical details on the calculation of high Magnus
orders in Refs\@.~\citep{Kuwahara2016b}. This research was supported
by a grant from the United States-Israel Binational Foundation (BSF,
Grant No. 2019644), Jerusalem, Israel, and by the Israel Science Foundation
(grants No. 527/19 and 218/19).
\end{acknowledgments}

\bibliography{local,lib_yevgeny}

\beginsupplement

\global\long\def\E{\mathrm{e}}%
\global\long\def\D{\mathrm{d}}%
\global\long\def\I{\mathrm{i}}%
\global\long\def\mat#1{\mathsf{#1}}%
\global\long\def\vec#1{\mathsf{#1}}%
\global\long\def\cf{\textit{cf.}}%
\global\long\def\ie{\textit{i.e.}}%
\global\long\def\eg{\textit{e.g.}}%
\global\long\def\vs{\textit{vs.}}%
 
\global\long\def\ket#1{\left|#1\right\rangle }%

\global\long\def\etal{\textit{et al.}}%
\global\long\def\tr{\text{Tr}\,}%
 
\global\long\def\im{\text{Im}\,}%
 
\global\long\def\re{\text{Re}\,}%
 
\global\long\def\bra#1{\left\langle #1\right|}%
 
\global\long\def\braket#1#2{\left.\left\langle #1\right|#2\right\rangle }%
 
\global\long\def\obracket#1#2#3{\left\langle #1\right|#2\left|#3\right\rangle }%
 
\global\long\def\proj#1#2{\left.\left.\left|#1\right\rangle \right\langle #2\right|}%
\global\long\def\mds#1{\mathds{#1}}%

\begin{center}
\textbf{\large{}Supplementary Material: Transport in Stark Many Body
Localized Systems}\textbf{\Large{}}\\
\par\end{center}

\section{Transition location}

\begin{figure}[H]
\centering{}\includegraphics{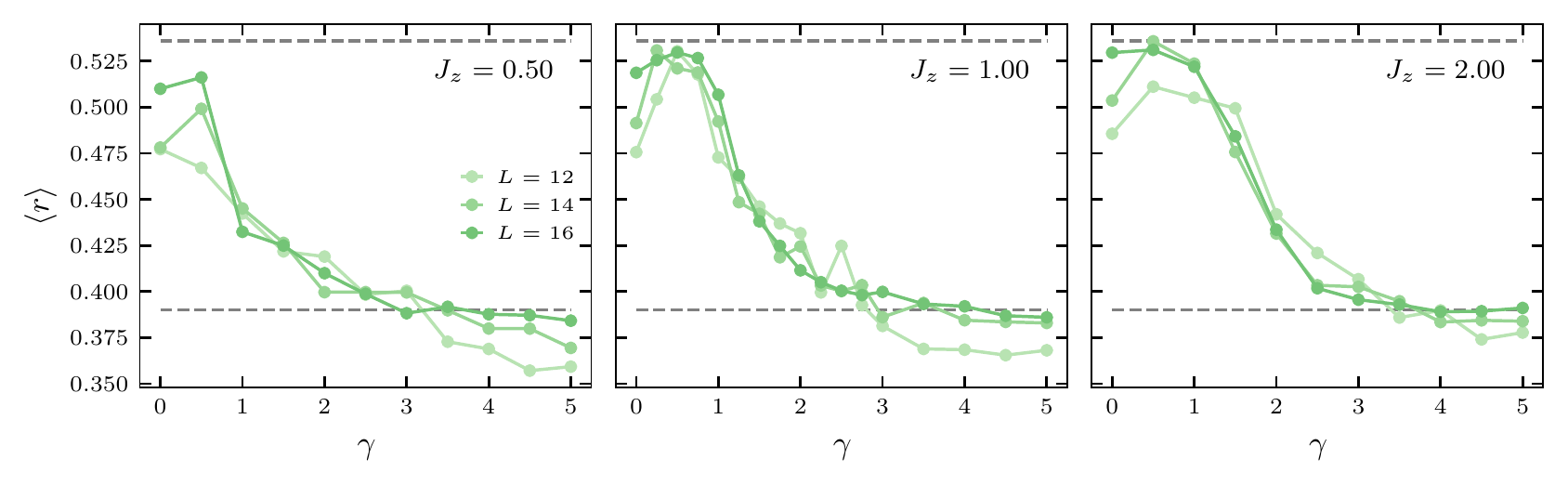}\caption{\label{fig:r-metric}$\left\langle r\right\rangle $ as of a function
of electric field strength for various interaction strengths (different
panels), and system sizes. Larger system size corresponds to stronger
color intensity. The black dashed lines correspond to WD statistics
($r\simeq0.536$) and Poisson ($r\simeq0.39$) statistics. The model
parameters that were used are $J_{xy}=2,\ J_{z}\in[0.5,1,2],\ \alpha=0.5$.}
\end{figure}
To approximately assess the location of the Stark-MBL transition we
use the standard metric,

\begin{equation}
r_{\alpha}=\text{min}\left(\frac{s_{\alpha}}{s_{\alpha-1}},\frac{s_{\alpha-1}}{s_{\alpha}}\right),
\end{equation}
where $s_{\alpha}\equiv E_{\alpha+1}-E_{\alpha}$ are the spacing
between adjacent eigenvalues of the Hamiltonian. For integrable systems
the mean of this quantity ($\left\langle r\right\rangle $), is typically
given by $\left\langle r\right\rangle \approx0.39$, which corresponds
to a Poissonian distribution, while for quantum chaotic systems it
is $\left\langle r\right\rangle \approx0.536$, which\lyxadded{Guy Zisling}{Mon Aug 16 10:14:21 2021}{
}corresponds to Wigner Dyson distribution. In Fig.~\ref{fig:r-metric}
we examine $\left\langle r\right\rangle $ as a function of the electric
field strength $\gamma$ for various couplings $J_{z}$. We observe
a transition from a Wigner-Dyson distribution for low electric fields
to a Poissonian distribution at high electric fields. The transition
occurs approximately at $\gamma\approx J_{z}$. This analysis does
not depend strongly on the size of the system, in contrast to the
mean-square displacement results presented in the main text. The middle
panel ($J_{z}=1$) is in agreement with Ref.~\citep{VanNieuwenburg2019}
although we have used a different mechanism to break the symmetries
of the model.

\section{Delocalization time extraction}

\begin{figure}[t]
\centering{}\includegraphics{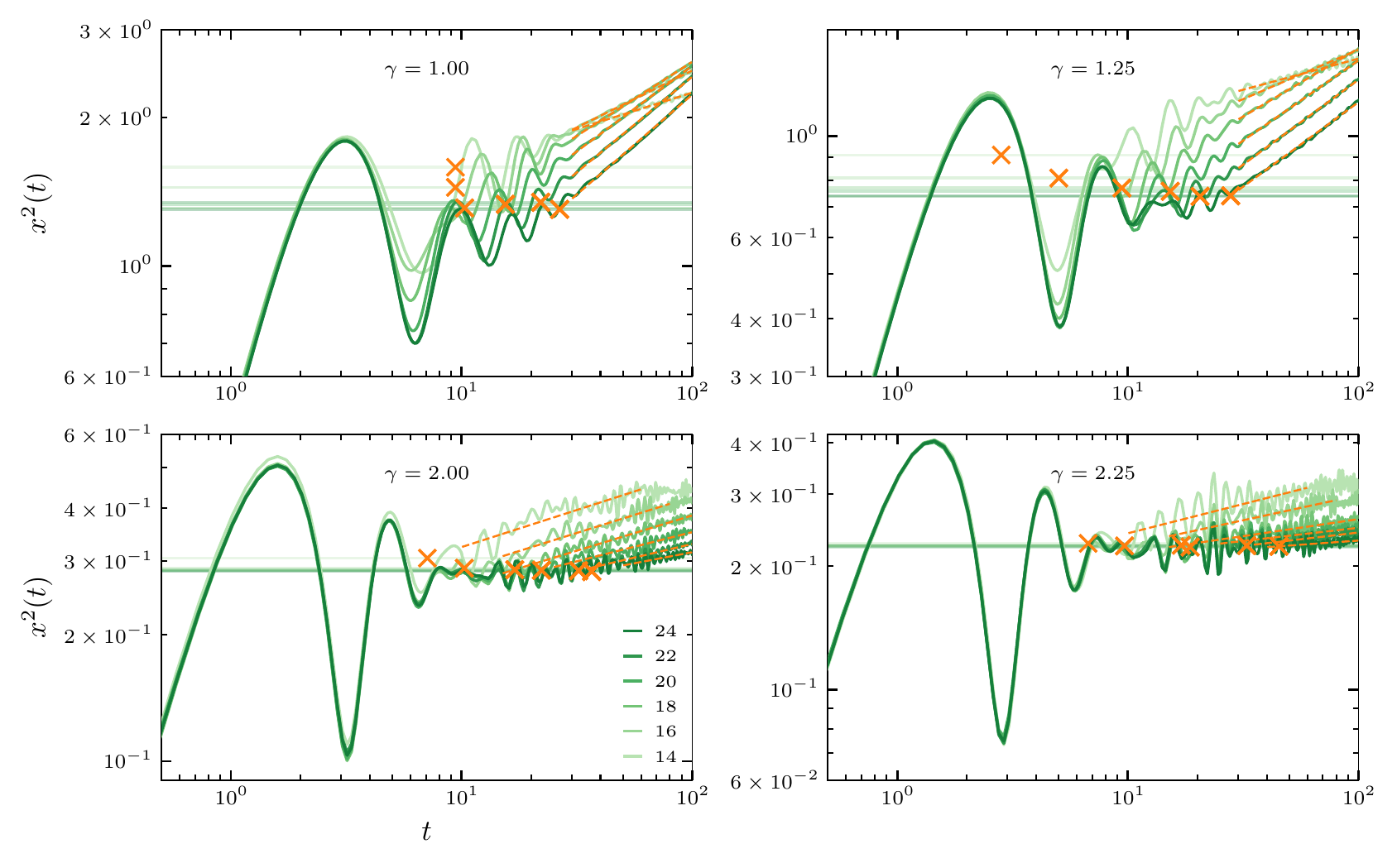}\caption{\label{fig:msd-static}Mean-square displacement (MSD) as a function
of time for $L\in[14,24]$ (Krylov based method). The orange dashed
line correspond to power-law fits ($x^{2}\sim t^{a}$), while the
horizontal lines indicate the plateau of the MSD calculated by taking
the mean of the MSD between the 2nd and the 3rd peaks. The orange
crosses are the estimated delocalization time $t^{\star}\left(\gamma,L\right)$
obtained from the intersection of the power-law fits with the plateau.
The color of the plateau lines matches the coloring of the corresponding
system size. All plots were obtained using $J_{xy}=2,\ J_{z}=1$.}
\end{figure}

In Fig.~\ref{fig:msd-static} we present the analysis used to obtain
$t^{\star}\left(\gamma,L\right)$ in Fig.~\ref{fig:t-star-scaling}
in the main text. The mean-square displacement (MSD) shows severe
finite size effects, with subdiffusive behavior delayed to later times
for larger system sizes. The locations of the plateaus (green horizontal
lines) are calculated by taking the mean of the MSD between the 2nd
and the 3rd peaks of the MSD. We fit the late time behavior with a
power-law fit, $x^{2}\propto t^{a}$ (orange dashed lines), and estimate
the delocalization time $t^{\star}\left(\gamma,L\right)$ by the intersection
of the plateaus with the power-law fits (orange crosses).

\section{Finite-size subdiffusive behavior}

\begin{figure}[th]
\includegraphics{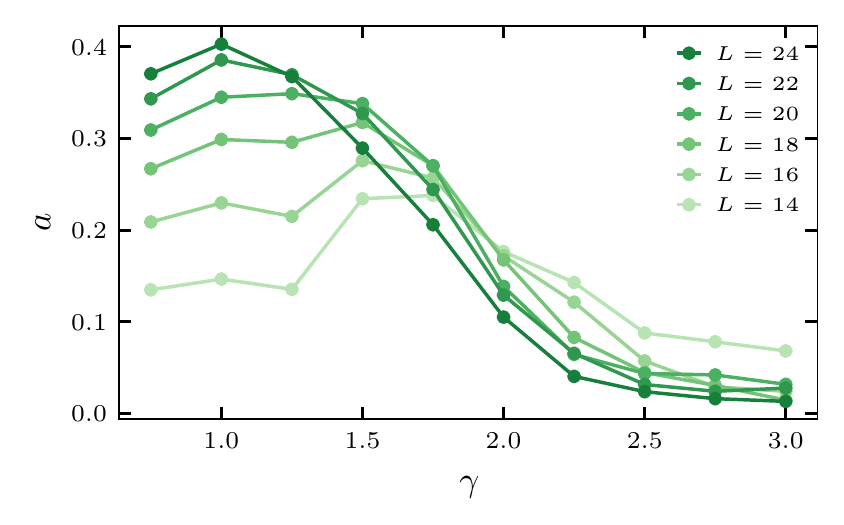}

\caption{\label{fig:diff_vs_gamma}The dynamical exponent $a$ as obtained
from the fits to the MSD, $x^{2}\propto t^{a}$ (see Fig.~\ref{fig:msd-static}),
as function of $\gamma$ for various system sizes ($L\in[14,24]$).}
\end{figure}

From the power law-fits in Fig.~\ref{fig:msd-static} we can obtain
the dynamical exponent $a$, which corresponds to the late-time growth
of the MSD, $x^{2}\sim t^{a}.$ We plot this exponent as a function
of the electric field $\gamma$ and for various system sizes in Fig.~\ref{fig:diff_vs_gamma}.
One can see an apparent transition between a subdiffusive behavior
($a<1$) to a localized behavior ($a\sim0$), with very strong finite-size
effects.While for $\gamma>J_{z}$ the exponent seems to converge with
the size of the system, it is important to keep in mind that the onset
of the subdiffusive transport is pushed to later times for larger
system sizes, as one can see in the main text and in Fig.~\ref{fig:msd-static}
indicating that the observed subdiffusive behavior is a finite-size
effect.

\section{Sensitivity to boundary conditions}

\begin{figure}[th]
\includegraphics{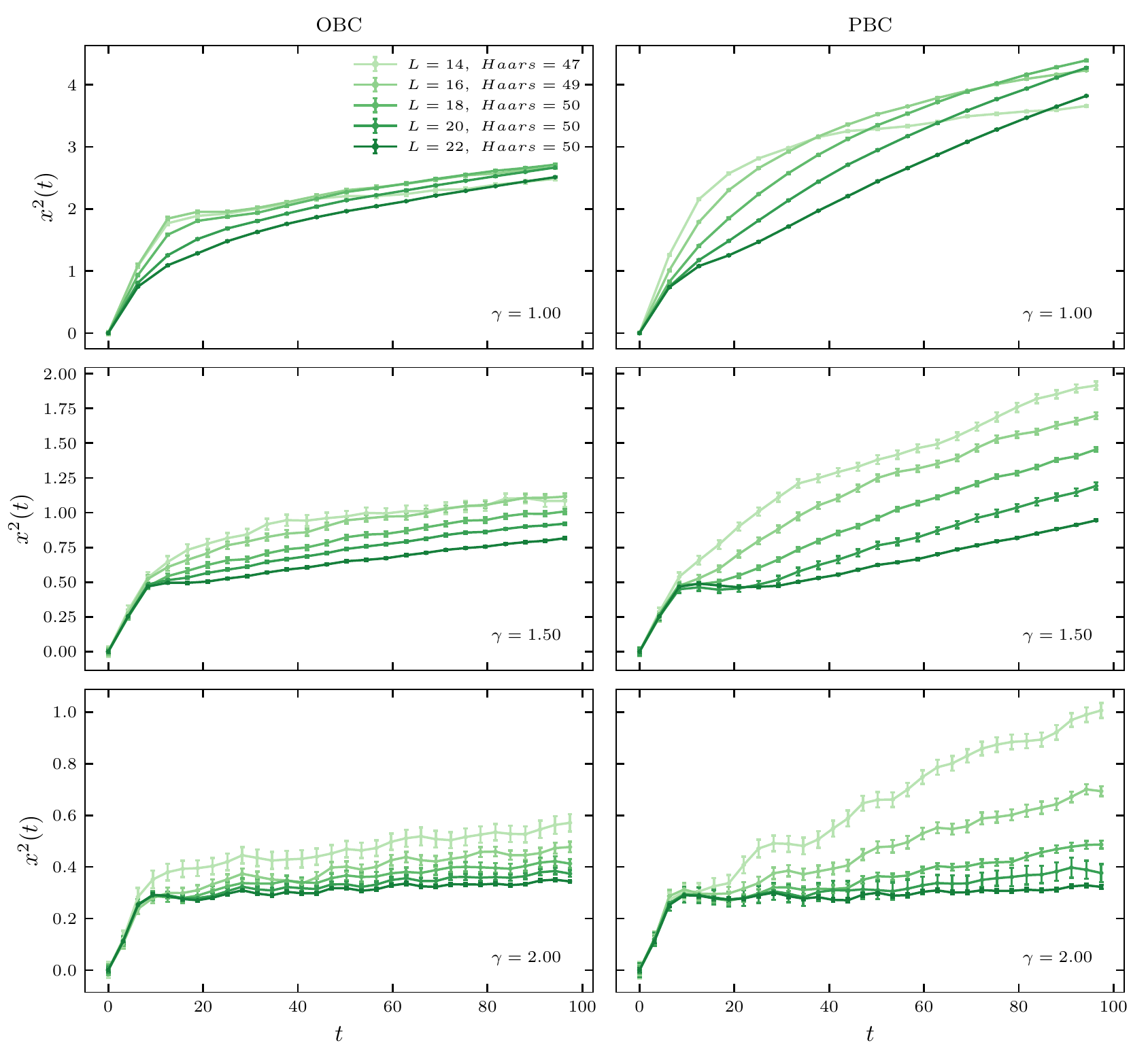}\caption{\label{fig:msd-dynamic}Mean-square displacement (MSD) as a function
of time for $L\in[14,24]$ (Krylov based method) calculated from the
dynamic gauge (\ref{eq:time-dep-ham}) in the main text. Left column:
open boundary conditions (OBC). Right column: periodic boundary conditions
(PBC). Different rows have different electric fields $\gamma\in[1,1.5,2]$.
All plots were obtained for $J_{xy}=2,\ J_{z}=1$.}
\end{figure}

In this Section we show that the conclusions of the main text are
robust to changes in the gauge and the boundary conditions. In Fig.~\ref{fig:msd-dynamic}
we have calculated the MSD as a function of time, using the dynamical
gauge, (\ref{eq:time-dep-ham}) in the main text for various electric
fields $\gamma$ (rows), various system sizes (color intensity), and
two different boundary conditions (columns). We see that in the dynamic
gauge the MSD shows less pronounced oscillations compared to the static
gauge, allowing to spot the formation of the localization plateau
already for $\gamma=1.$ The results remain qualitatively the same
to the results in the static gauge (Fig.\ref{fig:msd-static}), with
severe finite size effects, and a delocalization time that is increasing
with the system size. The quantitative difference between open and
periodic boundary conditions serves as another indication of finite-size
effects, though the localization plateau for both boundary conditions
appears at about the same MSD.

\section{Dynamical behavior of truncated effective Hamiltonians}

\begin{figure}[th]
\includegraphics{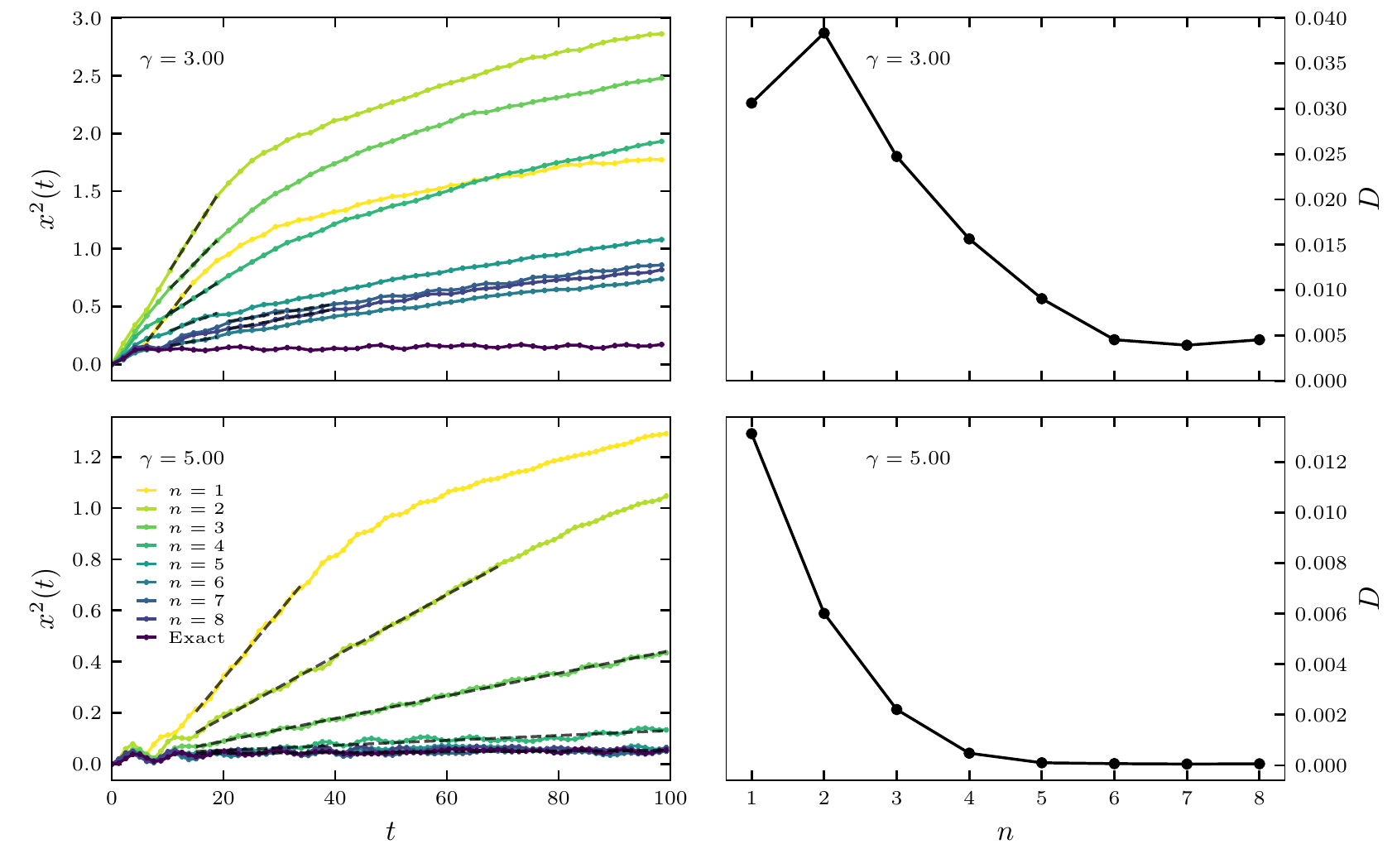}\caption{\label{fig:msd-magnus-diff}Mean-squared displacement as a function
of time for two electric fields (left column). The darkest lines correspond
to numerically exact results obtained using (Eq.~\ref{eq:time-dep-ham}
in the main text). The colored lines with increasing intensity corresponds
to evolution using effective Hamiltonians (\ref{eq:s_magnus-truncated}),
obtained from a truncated Magnus expansion. The black dashed lines
corresponds to linear fits, $x^{2}\sim2D\,t$, and the diffusion coefficient
$D$ is plotted in the right column as function of the truncation
order. For both $\gamma=3,5$ there is a visible trend of $D\propto1/n$.
The parameters used are, $J_{xy}=2,\ J_{z}=1,\ L=14$.}
\end{figure}

In this Section we study the late-times dynamical behavior of the
effective Hamiltonians calculated using Magnus expansion in $\gamma^{-1}$
up to some order $n$. In Fig.~\ref{fig:msd-magnus-diff} (left column)
we calculate the MSD for two electric fields (rows). We see that it
develops a pronounced linear behavior, indicative of diffusion, $x^{2}\sim2Dt$,
where $D$ is the linear response diffusion coefficient. For even
longer times the MSD saturates, since the system is finite. We extract
the diffusion coefficient from the relevant time windows (black dashed
lines in Fig.~\ref{fig:msd-magnus-diff}), and plot it as a function
of the truncation order, $n$ on the right column of Fig.~\ref{fig:msd-magnus-diff}.

The diffusion coefficient $D\left(\gamma,n\right)$ is monotonically
decreasing with the order of the Magnus expansion and the strength
of the electric field, approximately following $D\sim1/n$. While
this finding indicates that the truncated effective Hamiltonian is
delocalized, it doesn't imply much on the original interacting Stark
model, since the diffusive behavior of the effective Hamiltonian emerges
for at times for which the dynamics under the effective Hamiltonian
doesn't \emph{not} well approximate the numerically exact dynamics.
What is interesting, is that the infinite order Magnus expansion,
if it is convergent, could correspond to localized dynamics\lyxadded{Guy Zisling}{Mon Aug 16 10:24:23 2021}{}.

\section{Convergence criteria of the Magnus expansion}

\begin{figure}
\includegraphics{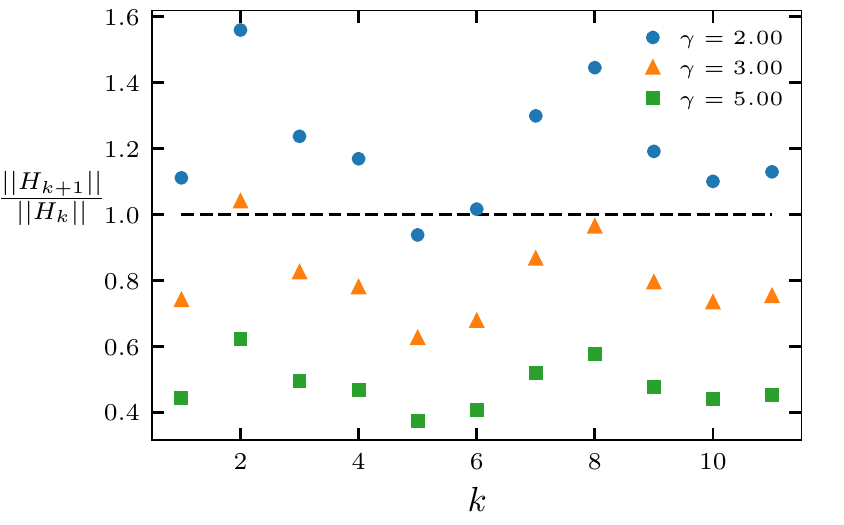}\caption{\label{fig:magnus_norms}D'Alembert criterion of convergence as a
function of the Magnus expansion order ($k$), $\left\Vert \hat{H}_{k+1}\right\Vert /\left\Vert \hat{H}_{k}\right\Vert <1$
(see (\ref{eq:s_magnus-truncated}) for definition). Different colors
(markers) represents different electric field strength $\gamma=2,3$
and 5. The dashed black line corresponds to a convergence requirement..
The parameters used are, $J_{xy}=2,\ J_{z}=1,\ L=10$.}
\end{figure}

In this Section we examine the convergence of the Magnus expansion
of the effective Hamiltonian,

\begin{equation}
\hat{H}_{\text{eff}}^{\left(n\right)}=\sum_{k=0}^{n}\hat{H}_{k},\label{eq:s_magnus-truncated}
\end{equation}
while each term $\hat{H}_{k}$ is of the order of $\gamma^{-k}$.
The D'Alembert criterion of convergence is $\left\Vert \hat{H}_{k+1}\right\Vert /\left\Vert \hat{H}_{k}\right\Vert <1$,
where $\left\Vert .\right\Vert $ indicates the operator norm. In
Fig.~\ref{fig:magnus_norms} we the D'Alembert criterion is presented
for different electric fields, $\gamma$. We see that while for $\gamma\leq2$
the series is divergent, for $\gamma\apprge3$ is it convergent at
least up to 10th order. We note that this doesn't necessarily mean
that the series has a finite radius of convergence, since divergence
can occur for relatively large expansions orders \citep{Kuwahara2016b}.

\section{Density of states of the truncated effective Hamiltonians}

\begin{figure}[th]
\includegraphics{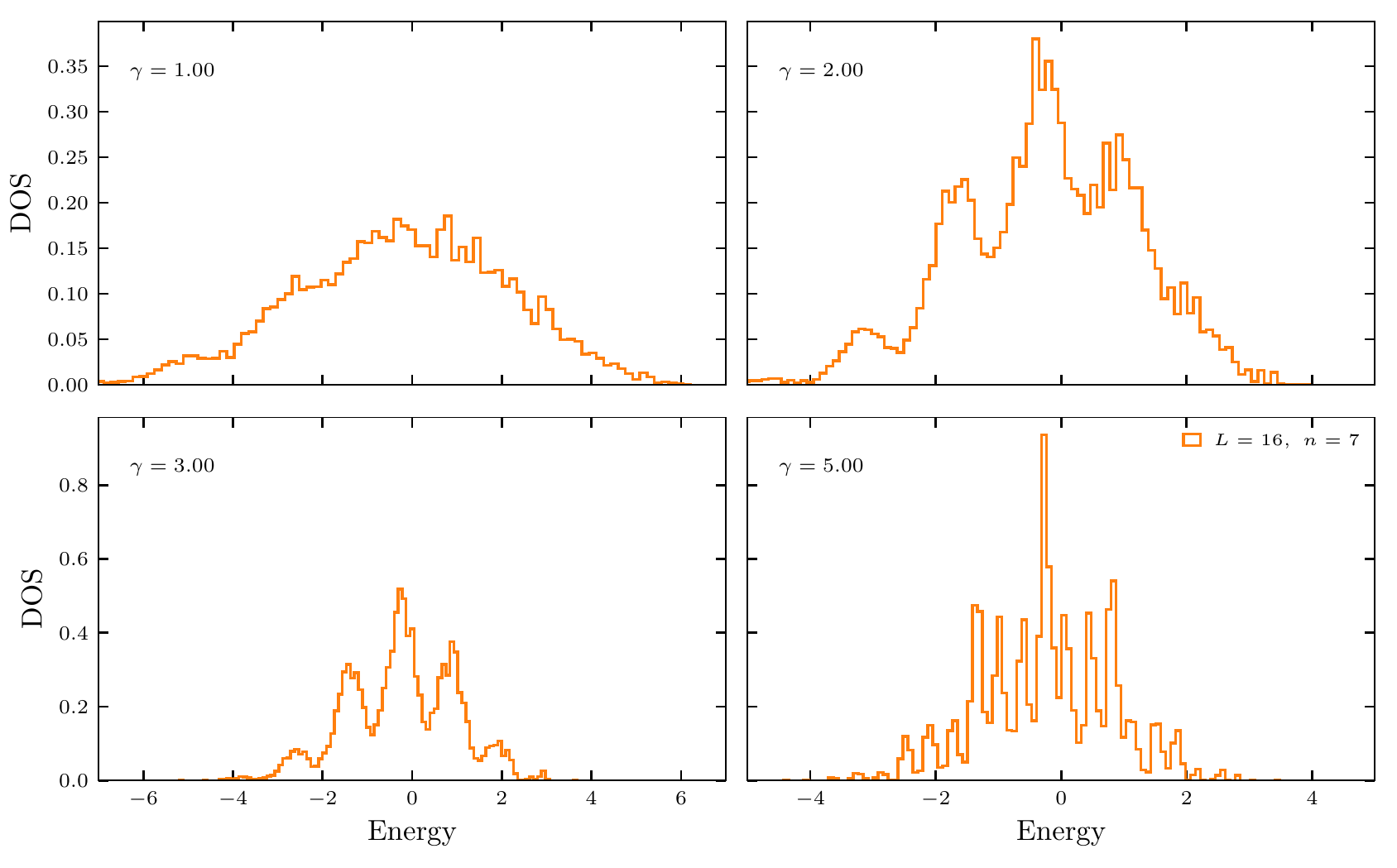}\caption{\label{fig:magnus_dos}The density of states of the truncated effective
Hamiltonian $\hat{H}_{\text{eff}}^{\left(n=7\right)}$ for $L=16$
and $\gamma=1,2,3$ and \lyxdeleted{Guy Zisling}{Mon Aug 16 10:27:30 2021}{5}\lyxadded{Guy Zisling}{Mon Aug 16 10:27:30 2021}{$5$}.
.All plots were obtained for $J_{xy}=2,\ J_{z}=1$.}
\end{figure}

The zero order truncated effective Hamiltonian, $\hat{H}_{\text{eff}}^{\left(0\right)}$
in (\ref{eq:s_magnus-truncated}) corresponds to the interaction term,
\begin{equation}
\hat{H}_{\text{eff}}^{\left(0\right)}=J_{z}\sum_{i=1}^{L}\hat{S}_{i}^{z}\hat{S}_{i+1}^{z},
\end{equation}
whose spectrum is composed of equally spaced degenerate bands, separated
$J_{z}/4$ apart. The following terms of the expansion are of order
$J_{z}/\gamma$, and they partially lift this degeneracy giving a
width of $J_{z}/\gamma$ to the bands. To demonstrate this in Fig.~\ref{fig:magnus_dos}
we plot the density of states (DOS) of $\hat{H}_{\text{eff}}^{\left(n\right)}$
for a number of electric fields, $\gamma$. While the gaps are washed
away for $J_{z}/4<J_{z}/\gamma$, namely $\gamma<4$, they become
clearly visible as $\gamma$ increases.

\section{Finite-size analysis}

\begin{figure}[th]
\centering{}\includegraphics{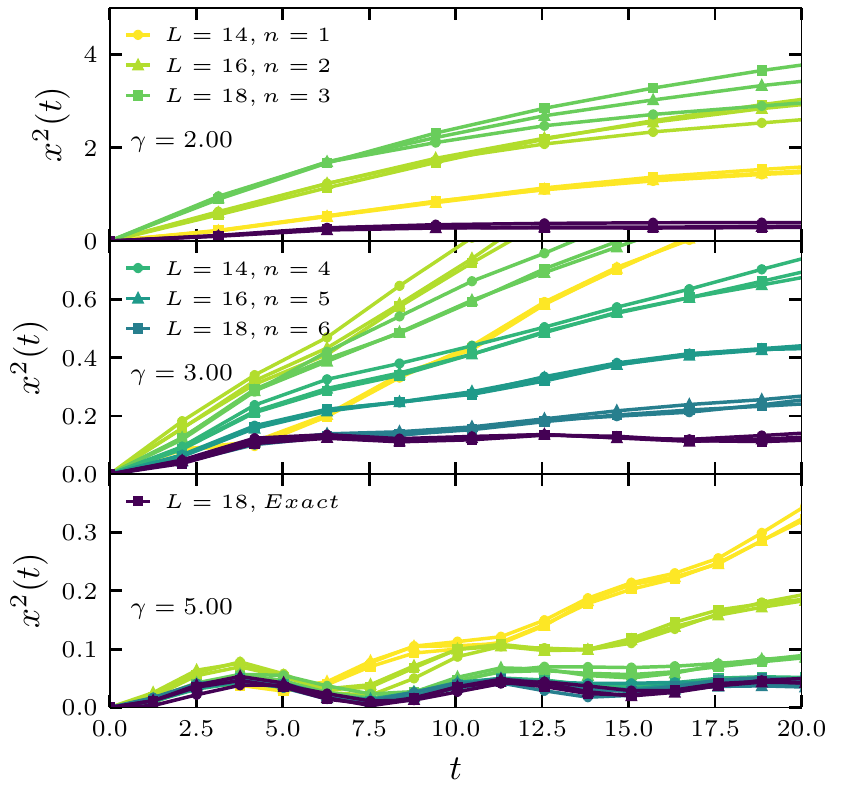}\caption{\label{fig:magnus-msd-fze}Finite size analysis of Fig.~\ref{fig:Magnus}
in the main text. Mean-squared displacement as a function of time
for various electric fields. The darkest lines correspond to numerically
exact results obtained using (Eq.~(\ref{eq:time-dep-ham}) in the
main text). The colored lines corresponds to an evolution using the
effective Hamiltonian (\ref{eq:s_magnus-truncated}), obtained from
a truncated Magnus expansion. Different markers ($\CIRCLE$, $\blacktriangle$,
$\blacksquare$)\lyxadded{Guy Zisling}{Mon Aug 16 10:25:52 2021}{}
stand for different system sizes $L=14,16$ and 18, correspondingly.
The used parameters are, $J_{xy}=2,\ J_{z}=1$.}
\end{figure}

In Fig.~\ref{fig:magnus-msd-fze} we repeat the analysis of Fig.~\ref{fig:Magnus}
from the main text for a number of system sizes, showing that there
are no considerable system size dependence in the determination of
$t_{\text{Magnus}}$, namely the time up to which there is a reasonable
agreement between the MSD computed using $\hat{H}_{\text{eff}}^{\left(n\right)}$
and the MSD of the dynamical gauge Hamiltonian (\ref{eq:time-dep-ham})
in the main text.

\end{document}